\documentclass[english]{revtex4-2}
\pdfoutput=1
\usepackage[T1]{fontenc}
\usepackage[utf8]{inputenc}
\usepackage{babel}
\usepackage{amsmath}
\usepackage{amssymb}
\usepackage{graphicx}
\usepackage{esint}
\usepackage[unicode=true,pdfusetitle,
 bookmarks=true,bookmarksnumbered=false,bookmarksopen=false,
 breaklinks=false,pdfborder={0 0 1},backref=false,colorlinks=false]
 {hyperref}

\makeatletter

\providecommand{\tabularnewline}{\\}

\makeatother

\begin{document}
\title{A multi particle toy system with analytic solutions to investigate
composite bosons in a harmonic potential}
\author{Detlef Schmicker}
\email{dschmicker@physik.de}

\date{\today}
\begin{abstract}
We construct a three dimensional toy systems with two types of fermions
forming a composite boson. They are hold in a harmonic potential.
The basis functions are constructed from an internal and an external
Gauss function. All integrals have analytical solutions. The high
symmetry reduces the number of integrals to be calculated for the
symmetrized wave functions. With the internal Gauss function the composite
bosons can be tuned from fermionic unbound behavior to bosonic bound
behavior. 
\end{abstract}
\maketitle

\section{Introduction}

In physics many particles handled as bosons are composed of fermions,
for example $^{4}He$. Of cause they can not be perfect bosons. An
established criteria is the entanglement between the fermions building
the boson \citep{quantum-entanglement}. Entanglement measures are
difficult to handle, especially if identical particles are involved,
which is always the case in multi particle systems \citep{entanglement-many-body}.
Taking this into account it might be handy to have a composite boson
toy system with several particles, which is computational handable,
to experiment with different particle numbers and entanglements. After
the necessary matrix elements are calculated, which is done (e. g.
with mathematica) within hours on a PC for up to 18 total particles,
all properties can be calculated from analytical formulas. We hold
the composite bosons in a 3D harmonic potential and will tune them
continuously from fermionic behavior of the consisting fermions to
bosonic behaviour of the composite bosons.

\section{Construction of the wave function}

A very simple composite boson is the hydrogen atom consisting of two
1/2 spin particles. To keep it simple our toy system will not have
a different mass for both particles. All particles will be localized
by a trap potential. A simple unsymmetrized three dimensional wave
function for n particles of each type (a and b) is
\begin{equation}
\psi(\overrightarrow{a}_{1},\overrightarrow{a}_{2},...,\overrightarrow{a}_{n},\overrightarrow{b}_{1},\overrightarrow{b}_{2},...,\overrightarrow{b}_{n})=\prod_{i=1}^{n}e^{-p\cdot(\overrightarrow{a}_{i}^{2}+\overrightarrow{b^{2}}_{i})}\cdot\prod_{i=1}^{n}e^{-q\cdot(\overrightarrow{a}_{i}-\overrightarrow{b}_{i})^{2}}\ with\ widths\ \frac{1}{\sqrt{p}}\ and\ \frac{1}{\sqrt{q}}\,.\label{eq:unsymmetrized wave function}
\end{equation}
This wave function has to be symmetrized or antisymmetrized with respect
to the different particles
\[
\phi(\overrightarrow{a}_{1},\overrightarrow{a}_{2},...,\overrightarrow{a}_{n},\overrightarrow{b}_{1},\overrightarrow{b}_{2},...,\overrightarrow{b}_{n})=\sum_{P_{i}(\overrightarrow{a}),P_{j}(\overrightarrow{b})}S(P_{i}(\overrightarrow{a}))\cdot S(P_{j}(\overrightarrow{b}))\cdot\psi(P_{i}(\overrightarrow{a}),P_{j}(\overrightarrow{b}))\,,
\]
with P indicating the permutations of the variables and S(P) indication
the signature of the permutation (or 1, if one wants to investigate
bosons for comparison).

\section{Matrix elements of the symmetrized wave function}

The overlap integral is calculated as
\[
O=\int_{\overrightarrow{a}_{i}\overrightarrow{b}_{i}}\phi(\overrightarrow{a}_{1},\overrightarrow{a}_{2},...,\overrightarrow{a}_{n},\overrightarrow{b}_{1},\overrightarrow{b}_{2},...,\overrightarrow{b}_{n})\cdot\phi(\overrightarrow{a}_{1},\overrightarrow{a}_{2},...,\overrightarrow{a}_{n},\overrightarrow{b}_{1},\overrightarrow{b}_{2},...,\overrightarrow{b}_{n})\ d\overrightarrow{a}_{1}d\overrightarrow{a}_{2}\cdot\cdot\cdot d\overrightarrow{a}_{n}\cdot d\overrightarrow{b}_{1}d\overrightarrow{b}_{2}\cdot\cdot\cdot d\overrightarrow{b}_{n}\,,
\]
which leads to a number of different summands of the $\psi$ functions.
Basically there is a sum
\begin{multline*}
O=\sum_{P_{i}(\overrightarrow{a}'),P_{j}(\overrightarrow{b}')}\sum_{P_{i}(\overrightarrow{a}),P_{j}(\overrightarrow{b})}S(P_{i}(\overrightarrow{a}'))\cdot S(P_{j}(\overrightarrow{b}'))S(P_{i}(\overrightarrow{a}))\cdot S(P_{j}(\overrightarrow{b}))\cdot\\
\left(\int_{a_{i}b_{i}}\psi(P_{i}(\overrightarrow{a}'),P_{j}(\overrightarrow{b}'))\cdot\psi(P_{i}(\overrightarrow{a}),P_{j}(\overrightarrow{b}))d^{n}\overrightarrow{a}\,d^{n}\overrightarrow{b}\right)\,,
\end{multline*}
and we just need the integrals in every summand. 
\[
\int_{\overrightarrow{a}_{i}\overrightarrow{b}_{i}}\psi(P_{i}(\overrightarrow{a}'),P_{j}(\overrightarrow{b}'))\cdot\psi(P_{i}(\overrightarrow{a}),P_{j}(\overrightarrow{b}))d^{n}\overrightarrow{a}\,d^{n}\overrightarrow{b}
\]
Due to symmetry the number of integrals is drastically reduced. The
same holds for the kinetic and potential energy terms which include
an Operator O

\[
\int_{\overrightarrow{a}_{i}\overrightarrow{b}_{i}}\psi(P_{i}(\overrightarrow{a}'),P_{j}(\overrightarrow{b}'))\cdot O_{kin,or,pot}\circ\psi(P_{i}(\overrightarrow{a}),P_{j}(\overrightarrow{b}))d^{n}\overrightarrow{a}\,d^{n}\overrightarrow{b}\,.
\]
Without loosing generality the overall ordering of a and b is free,
as all a particles and all b particles are identical

\[
\int_{\overrightarrow{a}_{i}\overrightarrow{b}_{i}}\psi(P_{i}(\overrightarrow{a}'),P_{j}(\overrightarrow{b}'))\cdot O_{kin,or,pot}\circ\psi((\overrightarrow{a}_{1},\overrightarrow{a}_{2},...,\overrightarrow{a}_{n},\overrightarrow{b}_{1},\overrightarrow{b}_{2},...,\overrightarrow{b}_{n})d^{n}\overrightarrow{a}\,d^{n}\overrightarrow{b}\,.
\]

The $\psi(P_{i}(\overrightarrow{a}'),P_{j}(\overrightarrow{b}'))$
are symmetric for exchanging $a_{i}$and $b_{i}$ with $a_{j}$and
$b_{j}$, as can be seen in equation \ref{eq:unsymmetrized wave function},
therefor it can be ordered

\[
\int_{\overrightarrow{a}_{i}\overrightarrow{b}_{i}}\psi(\overrightarrow{a}_{1},\overrightarrow{a}_{2},...,\overrightarrow{a}_{n},P(\overrightarrow{b}'))\cdot O_{kin,or,pot}\circ\psi(\overrightarrow{a}_{1},\overrightarrow{a}_{2},...,\overrightarrow{a}_{n},\overrightarrow{b}_{1},\overrightarrow{b}_{2},...,\overrightarrow{b}_{n})d^{n}\overrightarrow{a}\,d^{n}\overrightarrow{b}\,.
\]
Many of the permutations $P(b')$ are symmetric too. Basically one
can count the length of the chains which are coupling the positions
of the $b_{i}$ to the a due to the $e^{-q\cdot(a_{i}-b_{i})^{2}}$term.
If one wants to simplify the Laplacian in the kinetic energy term
to take only the particle $\overrightarrow{a}_{1}$ (and only one
direction) into account and multiplying the energy with the total
number of particle (and directions), as all particles are identical
for the kinetic energy, than one has to handle the length of the coupling
chain including $a_{1}$ separately. An example of a chains would
be

\[
\psi(\overrightarrow{a}_{1},\overrightarrow{a}_{2},\overrightarrow{a}_{3},\overrightarrow{a}_{4},\overrightarrow{b}_{3},\overrightarrow{b}_{1},\overrightarrow{b}_{2,}\overrightarrow{b}_{4})\rightarrow\overrightarrow{a}_{1}-\overrightarrow{b}_{3}-\overrightarrow{a}_{3}-\overrightarrow{b}_{2}-\overrightarrow{a}_{2}-\overrightarrow{b}_{1}-\overrightarrow{a}_{1}\ and\ \overrightarrow{a}_{4}-\overrightarrow{b}_{4}-\overrightarrow{a}_{4}\,.
\]
Taking this into account the number of matrix elements to be calculated
is shown in table \ref{tab:Number-of-matrix-elements}. The matrix
elements for 3 particles each are shown in table \ref{tab:Example-results-for-n3}.

\begin{table}
\begin{centering}
\begin{tabular}{|c|c|}
\hline 
number of particles of each kind & number of matrix elements to be calculated\tabularnewline
\hline 
\hline 
1 & 1\tabularnewline
\hline 
2 & 2\tabularnewline
\hline 
3 & 4\tabularnewline
\hline 
4 & 7\tabularnewline
\hline 
5 & 12\tabularnewline
\hline 
6 & 19\tabularnewline
\hline 
7 & 30\tabularnewline
\hline 
8 & 45\tabularnewline
\hline 
9 & 67\tabularnewline
\hline 
10 & 97\tabularnewline
\hline 
\end{tabular}
\par\end{centering}
\caption{Number of matrix elements for a 2n particle system\label{tab:Number-of-matrix-elements}}

\end{table}
\begin{table}
\begin{centering}
\begin{tabular}{|c|c|c|c|}
\hline 
Matrix Element & Factor & Overlap & Kinetic energy\tabularnewline
\hline 
\hline 
$\psi(\overrightarrow{a}_{1},\overrightarrow{a}_{2},\overrightarrow{a}_{3},\overrightarrow{b}_{1},\overrightarrow{b}_{2},\overrightarrow{b}_{3})$ & 1 & $\frac{\pi^{9}}{512(p(p+2q))^{9/2}}$ & $-\frac{\pi^{9}(p+q)}{512(p(p+2q))^{9/2}}$\tabularnewline
\hline 
$\psi(\overrightarrow{a}_{1},\overrightarrow{a}_{2},\overrightarrow{a}_{3},\overrightarrow{b}_{1},\overrightarrow{b}_{3},\overrightarrow{b}_{2})$ & -1 & $\frac{\pi^{9}}{512p^{3}(p+q)^{3}(p+2q)^{3}}$ & $-\frac{\pi^{9}}{512p^{3}(p+q)^{2}(p+2q)^{3}}$\tabularnewline
\hline 
$\psi(\overrightarrow{a}_{1},\overrightarrow{a}_{2},\overrightarrow{a}_{3},\overrightarrow{b}_{2},\overrightarrow{b}_{1},\overrightarrow{b}_{3})$ & -2 & $\frac{\pi^{9}}{512p^{3}(p+q)^{3}(p+2q)^{3}}$ & $-\frac{\pi^{9}\left(2p^{2}+4pq+q^{2}\right)}{1024p^{3}(p+q)^{4}(p+2q)^{3}}$\tabularnewline
\hline 
$\psi(\overrightarrow{a}_{1},\overrightarrow{a}_{2},\overrightarrow{a}_{3},\overrightarrow{b}_{2},\overrightarrow{b}_{3},\overrightarrow{b}_{1})$ & 2 & $\frac{\pi^{9}}{8(p(p+2q))^{3/2}\left(4p^{2}+8pq+3q^{2}\right)^{3}}$ & $-\frac{\pi^{9}\left(4p^{3}+12p^{2}q+9pq^{2}+q^{3}\right)}{8(p(p+2q))^{3/2}\left(4p^{2}+8pq+3q^{2}\right)^{4}}$\tabularnewline
\hline 
\end{tabular}\caption{Example results for 3 particles each\label{tab:Example-results-for-n3}}
\par\end{centering}
\centering{}The factor indicates how often the matrix element is found
in the sum and its sign is the signature of the permutation.
\end{table}

\section{The kinetic and potential energy}

For the kinetic energy the Operator $O_{kin}$ would be the Laplacian
$\nabla_{\overrightarrow{a_{1}},\overrightarrow{a_{2}},...,\overrightarrow{a_{n}},\overrightarrow{b_{1}},\overrightarrow{b_{2}},...,\overrightarrow{b_{n}}}$.
As due to symmetry the kinetic energy is the same for every particle
and every direction we can calculate the kinetic energy for the n
particle system with $O_{kin}=n\cdot\frac{d^{2}}{da_{11}^{2}}$, as
we will not include any constants in the Schrödinger equation in the
general calculations, especially we restrict our self to particles
$a$ and $b$ with the same mass $m=1/2$ as well as $\hbar=1$. 

As the potential energy is symmetric for all exchanges of identical
particles the number of matrix elements is the same as for the overlap
integral. After expansion all integrals are of the type
\[
\intop_{x=-\infty}^{\infty}e^{-ax^{2}+bx+c}\left(dx^{2}+ex+f\right)dx=\frac{\sqrt{\pi}e^{\frac{b^{2}}{4a}+c}\left(4a^{2}f+2a(be+d)+b^{2}d\right)}{4a^{5/2}}\,,
\]
which speeds up the integration significantly, leaving only simplifications
to be time consuming.

\section{Results for tunable composite bosons in a harmonic potential}

We use a harmonic potential in three dimensions without interactions
between the different fermions
\[
V_{pot}(\overrightarrow{a}_{1},\overrightarrow{a}_{2},...,\overrightarrow{a}_{n},\overrightarrow{b}_{1},\overrightarrow{b}_{2},...,\overrightarrow{b}_{n})=\sum_{i}\overrightarrow{a}_{i}^{2}+\overrightarrow{b}_{i}^{2}\,,
\]
together with a wave function (\ref{eq:unsymmetrized wave function})
which allows to tune the correlation of the composite bosons by the
parameter $q$ and $m=1/2$, as well as $\hbar=1$. The forming of
the composite bosons is controlled by the internal width $1/\sqrt{q}$,
which in a real physical system is usually a result of an interaction.
But as we want to keep it simple we just force the composite bosons
to have an internal width. A sample result for $1/\sqrt{q}=1$ is
shown in table (\ref{tab:result-for-harmonic}), where the external
energy per composite boson is calculated by substracting the internal
kinetic energy controlled the parameter $q$. The internal energy
is zero for $q=0$.

\begin{table}
\begin{centering}
\begin{tabular}{|c|c|c|c|c|}
\hline 
number of particles & Energy & width $1/\sqrt{p}$ & Energy per composite boson & External energy per composite boson\tabularnewline
\hline 
\hline 
1 & 9.37500 & 2.00000 & 9.37500 & 3.37500\tabularnewline
\hline 
2 & 21.8007 & 2.38716 & 10.9003 & 4.90033\tabularnewline
\hline 
3 & 35.2006 & 2.49827 & 11.7335 & 5.73354\tabularnewline
\hline 
4 & 49.5672 & 2.58060 & 12.3918 & 6.39179\tabularnewline
\hline 
5 & 64.8028 & 2.65889 & 12.9606 & 6.96056\tabularnewline
\hline 
6 & 80.7640 & 2.72260 & 13.4607 & 7.46066\tabularnewline
\hline 
7 & 97.3570 & 2.77376 & 13.9081 & 7.90814\tabularnewline
\hline 
8 & 114.522 & 2.81637 & 14.3153 & 8.31530\tabularnewline
\hline 
\end{tabular}
\par\end{centering}
\caption{Result for $1/\sqrt{q}=1$\label{tab:result-for-harmonic}}
Calculation of the energy for different numbers of composite bosons.
The external energy per composite boson is calculated by subtracting
the internal kinetic energy controlled by the internal parameter $q$
for one particle. The external energy of the one composite boson system
is 6 in this case (calculated e.g. with $q=0$). The internal width
was chosen to be $1/\sqrt{q}=1$ and the external width $1/\sqrt{p}$
was optimized to minimize the energy.
\end{table}

\begin{figure}
\begin{centering}
\includegraphics[width=10cm]{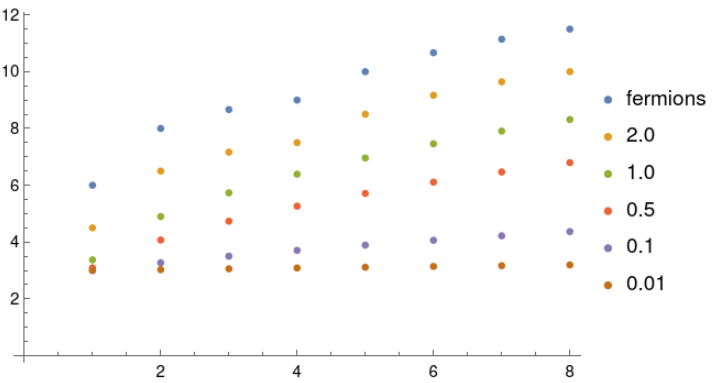}
\par\end{centering}
\begin{centering}
\caption{Composite bosons in harmonic potential for different coupling\label{fig:Composite-bosons-in-harmonic-potential}}
For internal widths $1/q^{2}$ of the composite bosons the external
energy per composite boson is shown depending on the number of composite
bosons. If they were perfect bosons the energy would be constant,
if the fermions building the bosons were not coupled at all the value
for ``fermions'' in an harmonic potential is added. With the parameter
$1/q^{2}$ it is possible to tune the composite bosons from bosonic
to fermionic character.
\par\end{centering}
\end{figure}

As we restricted our self to $m=1/2$ for the particle masses, we
will compare this to a quantum harmonic oscillator with the same harmonic
potential for both fermion types
\[
H=\sum_{i=1}^{3}\frac{p_{i}^{2}}{2m}+\frac{1}{2}m\omega^{2}x_{i}^{2}=\sum_{i=1}^{3}p_{i}+\frac{1}{4}\omega^{2}x_{i}^{2}\ ,
\]
with $\omega=2$. The energy states are
\[
E=\omega(n_{1}+n_{2}+n_{3}+\frac{3}{2})=2(n_{1}+n_{2}+n_{3}+\frac{3}{2})\ ,
\]
for each fermion type, corresponding to the external energy 6 in the
ground state with $n_{1}=n_{2}=n_{3}=0$. In the limit of boson we
do have $m=1$ and double force resulting in $E=3$ .

\section{Short Discussion and Outlook}

\begin{figure}
\begin{centering}
\includegraphics[width=10cm]{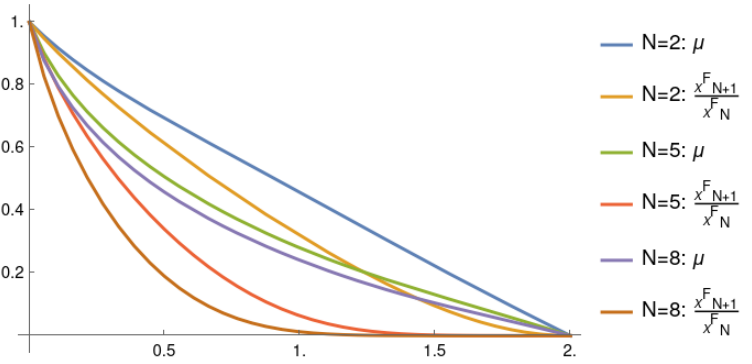}
\par\end{centering}
\begin{centering}
\caption{Comparison of the behavior of the c anhilation operator on $|N>$
with the fermion boson mixing parameter $\mu$\label{fig:Comparison-of-the}}
\par\end{centering}
\begin{centering}
Equation (23) in \citep{quantum-entanglement} compared to $\mu$
in (\ref{eq:Mixing-boson-fermion-mu})
\par\end{centering}
\centering{}
\end{figure}

The composite bosons can be tuned from fermionic to bosonic behavior
\begin{equation}
E_{composite}=\mu E_{boson}+(1-\mu)E_{fermion}\ ,\label{eq:Mixing-boson-fermion-mu}
\end{equation}
with $\mu$=0 in the fermionic case. They become better bosons ($\mu\rightarrow1$),
if their internal width $1/\sqrt{q}$ gets smaller. This $\mu$ might
be used as a parameter indicating how bosonic the behavior of the
system is. In figure (\ref{fig:Comparison-of-the}) we compare it
to the behavior of the c anhilation operator on the $|N>$ states
from \citep{quantum-entanglement}.

\begin{figure}
\begin{centering}
\includegraphics[width=8cm]{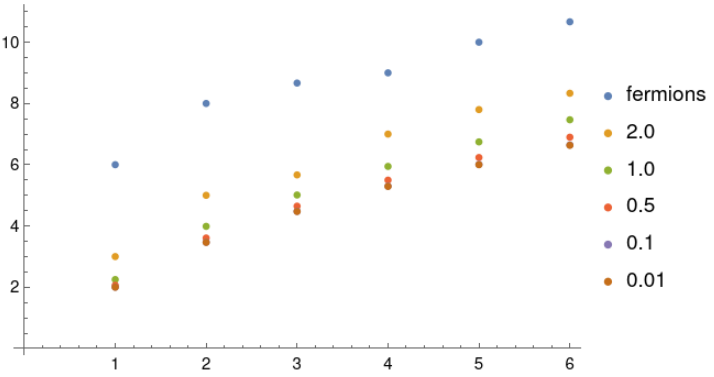}\hspace{1cm}\includegraphics[width=8cm]{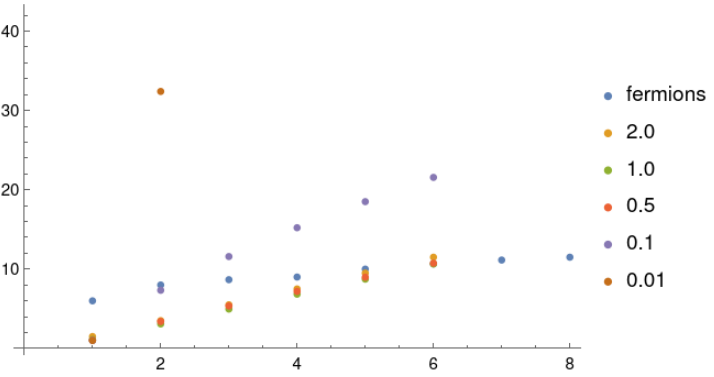}
\par\end{centering}
\caption{Composite bosons in harmonic potential for different coupling in lower
dimensions (2D 1D)\label{fig:Composite-bosons-in-low-dimension}}

\centering{}One can see, that in 2D (left) and 1D (right) the behavior
is not boson like \citep{CompositeBosonsLowDimensions}.
\end{figure}

Composite bosons depend strongly on the dimension \citep{CompositeBosonsLowDimensions}.
So we calculate fig (\ref{fig:Composite-bosons-in-harmonic-potential})
for lower dimensions in fig (\ref{fig:Composite-bosons-in-low-dimension}).
In both cases of lower dimensionthe bosonic behavior is not shown.
This effects are discussed in \citep{CompositeBosonsLowDimensions}.
It seems that the handling in \citep{quantum-entanglement} are hiding
some of the complexity of composite bosons, as no energies are involved
in the analysis. This might be a motivation for using a toy system
to quickly try different composite boson systems. 

\bibliographystyle{unsrt}
\bibliography{Literatur}

\end{document}